\documentstyle[12pt]{article}
\begin{document}

\input{epsf}

\def\beq{\begin{equation}}
\def\eq{\end{equation}}
\def\beqa{\begin{eqnarray}}
\def\eqa{\end{eqnarray}}
\def\bars{\begin{eqnarray*}}
\def\ears{\end{eqnarray*}}
\def\ov{\overline}
\def\ot{\otimes}

\def\deb{\frac{1}{2}}
\newcommand{\no}{\noindent}
\newcommand{\be}{\begin{equation}}
\newcommand{\ee}{\end{equation}}
\newcommand{\lesssim}{\raisebox{-0.5ex}{$\stackrel{<}{\sim}$}}
\newcommand{\dd}{\mbox{$\Delta$}}
\newcommand{\af}{\mbox{$\alpha$}}
\newcommand{\la}{\mbox{$\lambda$}}
\newcommand{\gh}{\mbox{$\gamma$}}
\newcommand{\ep}{\mbox{$\epsilon$}}
\newcommand{\vep}{\mbox{$\varepsilon$}}
\newcommand{\de}{\mbox{$\frac{1}{2}$}}
\newcommand{\tr}{\mbox{$\frac{1}{3}$}}
\newcommand{\qa}{\mbox{$\frac{1}{4}$}}
\newcommand{\sx}{\mbox{$\frac{1}{6}$}}
\newcommand{\vg}{\mbox{$\frac{1}{24}$}}
\newcommand{\tde}{\mbox{$\frac{3}{2}$}}
\newcommand{\lqcd}{\mbox{$\Lambda_{QCD}$}}
\newcommand{\xbj}{\mbox{$x_{bj}$}}
\newcommand{\jb}{\mbox{$\bar{j}$}}
\newcommand{\qb}{\mbox{$\bar{q}$}}

\newcommand{\np}{{\it Nucl. Phys.}}
\newcommand{\pl}{{\it Phys. Lett.}}
\newcommand{\prl}{{\it Phys. Rev. Lett.}}
\newcommand{\cmp}{{\it Commun. Math. Phys.}}
\newcommand{\jmp}{{\it J. Math. Phys.}}
\newcommand{\jpamg}{{\it J. Phys. {\bf A}: Math. Gen.}}
\newcommand{\lmp}{{\it Lett. Math. Phys.}}
\newcommand{\ptp}{{\it Prog. Theor. Phys.}}

\title{On the origin of the rise of $F_2$ at small $x$}
\author{{ H. Navelet$^1$}\\
{ R. Peschanski$^1$}\\
{ Ch. Royon$^2$}\\
{ L. Schoeffel$^2$}\\
{ S. Wallon$^1$}\\
 \\
{\small 1- Service de Physique Th\'eorique, Centre d'Etudes de Saclay}\\
{\small F-91191 Gif-sur-Yvette  Cedex, FRANCE}\\
{\small 2- DAPNIA - SPP, Centre d'Etudes de Saclay}\\
{\small F-91191 Gif-sur-Yvette  Cedex, FRANCE}}

\date{}
\maketitle

\begin{abstract}
\date{}

We show that, provided that the non-perturbative input
is regular at the right of the $\omega=0$  singularity of the dominant DGLAP anomalous dimension, the rise of $F_2$ at small $x,$  experimentally measured by the averaged
observable $\lambda = \left<\frac{\partial \ln F_2}{\partial \ln \frac{1}{x}} \right>,$ is input-independent in the perturbative $Q^2$ regime at small $x$. $\frac{\partial \ln xF}{\partial \ln Q^2}$ appears to be more input-dependent in the same range. The GRV-type parametrisations verify these properties.
Other models, namely the BFKL kernel(QCD dipoles), DGLAP(hard pomeron singularity) give different predictions for $\lambda$. At moderate $Q^2,$ there is a possibility of distinguishing these different perturbative QCD predictions in the near future.  
\\

\end{abstract}
\vspace{4cm}
\noindent
\hspace{1cm} August 1996\hfill\\
\hspace*{1cm} T96/094\hfill\\
\thispagestyle{empty}
\newpage
\setcounter{page}{1}

\no
I - The recently published 1994 results from HERA experiments 
on the proton structure function $F_2$ have reached a high level of
precision. It covers an extended kinematical range. In
particular, it reaches very low values of $x$ ($x \sim 10^{-5}$) and $Q^2$ ($Q^2 \sim 1.5 \, GeV^2$). These data confirm with high statistics the strong rise of $F_2$ when $x$ becomes very small, first noticed in 1992 experiments \cite{h2}. It is also observed in the newly reached kinematical range at small $Q^2$. This rise has been quantified \cite{h1} by a study of the observable 
\beq
\label{lambda}
\lambda = \left<\frac{\partial \ln F_2}{\partial \ln 1/x} \right>,
\eq
where the brackets mean that $\lambda$ is obtained from a fit of the form $F_2 \sim x^{-\lambda}$ at fixed  $Q^2$ and small values of $x$ ($x \leq .1$).
This observable has been previously proposed \cite{npw} as an interesting tool for discussing the various perturbative QCD expansions for the rise of $F_2$. 

Two kinds of perturbative QCD predictions are available to explain this rise.
At small $x$, the BFKL dynamics \cite{bfkl} naturally applies. It corresponds to the
multi-Regge regime of perturbative QCD. In this approach,
one sums up contributions of the type  $ \left(\alpha_ s {\rm ln}{1 \over
x} \right)^n $ . 
In the present paper,
we implement the BFKL dynamics using the recently developped QCD dipole model \cite{mueller}. It is based on the calculation of the infinite momentum wavefunction for arbitrary numbers of soft gluons in a heavy quark-antiquark (onium) state.
Combined with $k_T$-factorization, this framework has been successfully applied to proton
structure functions \cite{npr,nprw}.

In another approach, the well-known DGLAP evolution equations \cite{dglap} lead to alternative explanations for the rise of $F_2$, based on the renormalization group evolution and the operator product expansion. These equations can then be expressed in term of moments in the $\omega$-Mellin space.  In this approach, the matrix elements of the local operators can be either regular at the right of $\omega=0$ or singular in the $\omega$-Mellin space. The first class of models was illustrated in the paper \cite{wil} where a non-Regge behaviour for structure functions at small $x$ was first suggested. In a similar framework, a perturbative evolution of valence-like input 
distributions  lead to the  parametrization of Gl\"uck, Reya, Vogt (GRV) of structure functions \cite{grv}, which gave satisfactory predictions for HERA. The second possibility, initiated some years ago and recently revived by  L\'opez, Barreiro, Yndur\'ain  (LBY), uses a singular input and also provides  a satisfactory description of $F_2$\cite{LBY}. 

Our aim is to study the properties of $\lambda$ (formula (\ref{lambda})) namely its dependence on the perturbative QCD origin of the rise of $F_2$. The main results of this analysis are:

\noindent
i) In the GRV type parametrization, $\lambda$ is independent of the non-perturbative input for $x \leq 5.10^{-3}$ and uniquely determined by the DGLAP kernel singularity.

\noindent
ii) The different types of perturbative QCD predictions, BFKL (dipole), DGLAP (GRV), DGLAP (LBY), are compared to the data on $\lambda$.
They are compatible with
them but lead to significant differences  at moderate $Q^2$ $(1 \leq Q^2 \leq 10 \, GeV^2)$. This motivates precise measurements of $\lambda$
in this region.
\\
\eject
\no
II - Let us determine the predictions for $\lambda$.
First, we consider the direct and inverse Mellin transforms of the singlet structure function $xxF_s,$ namely
\beqa
\label{mellind}
F_s(\omega,Q^2,\mu^2) = \int^1_0 dx \, x^{\omega-1} xF_s(x,Q^2,\mu^2) \\
\label{mellini}
xF_s(x,Q^2,\mu^2)={1 \over 2i\pi} \int^{\omega_0 + i \infty}_{\omega_0 - i \infty}  x^{-\omega}
F_{s}(\omega,Q^2,\mu^2)  {\rm d} \omega,
\eqa
\no
where
the integration line $Re \omega=\omega_0 $ is
at the right of all singularities of $ xF_s(\omega,Q^2,\mu^2). $

\no
In the DGLAP scheme  $xF_s(\omega,Q^2,\mu^2)$ and $F_G(\omega,Q^2,\mu^2)$ (the gluon distribution in Mellin-moment space) verify
\beq
\label{dglapscheme}
\left(\begin{array}{c}
F_s(\omega,Q^2,\mu^2) \\
F_G(\omega,Q^2,\mu^2)
\end{array}\right) =
 K(\omega,Q^2,\mu^2)
\left(\begin{array}{c}
F_s(\omega,\mu^2) \\
F_G(\omega,\mu^2)
\end{array}\right)
\eq
\no
where $K(\omega,Q^2,\mu^2)$ is the Mellin transform of the DGLAP matrix kernel 
at the scale $Q^2$ ; The rightmost singularity of this kernel lies at
$\omega_0 = 0$. Also $F_{s}(\omega,\mu^2)$ is the Mellin transform of the singlet input which is either
regular or singular at the right of $\omega_0 = 0$, $F_G(\omega,\mu^2)$ is the Mellin transform of the gluon input. 
We then get two classes  of models satisfying a DGLAP evolution \cite{KLE}:

\no 
(i) Either the rightmost $\omega$-plane singularity is fixed by the DGLAP kernel singularity at $\omega = 0$, we obtain the GRV type parametrization.

\no
(ii) Or we have the  LBY type parametrization where the rightmost singularity lies at the right of $\omega = 0$ due to a singular input $F_s(\omega,Q^2,\mu^2)$, moreover its location is essentially not modified by the perturbative evolution.

\no
Let us observe that the BFKL dynamics also leads to a rightmost singularity at $\omega$ greater than $\omega = 0$. We will come back to this later.
\\

\no
{\bf a} -In the first case, we shall prove that the large $Q^2$ behaviour of $ \frac{\partial \ln xF_s}{\partial \ln \frac{1}{x}} $ is input independent and thus depends only on the DGLAP kernel.
Let us start with a valence-like input, that is for which the $\omega=0$  moment is well defined. In this case, the dominant $\omega-$plane singularity is generated by the DGLAP evolution equations \cite{dglap}  
\beqa
\label{ftilde}
F_s(\omega,Q^2,\mu^2) &=& \left({\nu_ F-\nu_ - \over \nu_+-\nu_ -} 
{\rm
exp} \ \nu_ +\xi +{\nu_ +-\nu_ F \over \nu_ +-\nu_- } {\rm exp} \ \nu_ -\xi
\right) \left(q+\bar q \right)(\omega) \nonumber \\
 &+&{2N_f\phi^ F_G \over \nu_ +-\nu_
F} \left( {\rm exp} \ \nu_ +\xi - {\rm exp} \ \nu_ -\xi \right)g(\omega) \nonumber \\ 
\eqa
\no
where 
\beq
\label{xidef}
\xi(\mu^2) = {\frac{1}{11 - {\frac{2}{3}} N_f}} \ln \left(\frac{{\ln}Q^2/\Lambda^2}{\ln  \mu^2/\Lambda^2} \right)
\eq
$q,\, \qb,\, g$ are the valence-like input at the low scale $Q^2 = \mu^2$.
$\nu_+,\, \nu_-,\, \Phi_G^F$ are the first-order DGLAP kernels \cite{dok}. $\Lambda$ is the one-loop QCD scale and $N_f$
is the number of active flavours.
\no
The assymptotic form of $\nu_{+}(\omega)$ near the $\omega=0$ singularity is   
\beq
\nu_{+}(\omega) \simeq \frac{4N_c}{\omega}-a
\eq
where $a$ is a constant ($a = \frac {11 N_c}{3} + \frac {2 N_f}{3 N_c^2} \sim 110/9$ for 3 active flavors).
Assymptotically, at a given $Q^2$  near the $\omega=0$ singularity we have 
\beq
F_s(\omega,Q^2,\mu^2) 
\simeq \omega  f(\omega) \exp {(\frac{4 N_c} {\omega} -a)\xi}
\eq
\no
where $f(\omega)$ is an input-dependent function regular at $\omega = 0$ by hypothesis. Hence this fonction can be expanded as
\beq
\label{deff}
f(\omega) = f(0) [1+\Sigma_{i} \omega^i  b_i]
\eq
\no
Then, we can find an expression for $xF_s(x,Q^2,\mu^2)$
\beq
xF_s(x,Q^2,\mu^2)={1 \over 2i\pi} \int^{\omega_0 + i \infty}_{\omega_0 - i \infty} 
\omega f(\omega) {\rm e}^{\  \omega {\ln}1/x} {\rm e}^{\frac{4 N_c} {\omega} \xi}  {\rm d} \omega
\eq

We now want to  demonstrate that $ \frac{\partial \ln xF_s}{\partial \ln \frac{1}{x}}$  is independent of the valence-like input. First, we can obtain an exact determination of this derivative when using the following property of Bessel functions 
\beq
{1 \over 2i\pi}\int^{\omega_0 + i \infty}_{\omega_0 - i \infty}
\omega^n {\rm e}^{\  \omega {\ln}1/x} {\rm e}^{\frac{4 N_c} {\omega} \xi} 
{\rm d} \omega = {(\frac{4 N_c \xi}{ \ln 1/x})}^{\frac{n+1}{2}} I_n[2{(4 N_c \xi \ln 1/x)}^{\frac{1}{2}}]
\eq
Inserting the form (9) into (10), it is possible to compute $xF_s(x,Q^2,\mu^2)$ as was already done in \cite{dok} for the lowest order in $\omega$.
Defining $\bar{\omega}$ and $v$ 
\beq
\bar{\omega} = {(\frac{4 N_c \xi}{ \ln 1/x})}^{\frac{1}{2}}
\eq
\beq
v = 2 {(4 N_c \xi \ln 1/x)}^{\frac{1}{2}},
\eq
\no
 it is straightforward to see that
\beq
\frac{{\partial \ln} xF_s } {{\partial \ln}{1 \over {x}}}
=
\bar{\omega} \frac{I_3(v)}{I_2(v)} \left[ 
\frac{1+\Sigma_i b_i {\bar{\omega}}^{i} \frac{I_{i+3}}{I_3} (v)}
{1+\Sigma_i b_i {\bar{\omega}}^{i} \frac{I_{i+2}}{I_2} (v)}\right]
\eq
At this stage we can see that the only dependence on the input comes from the different $b_{i}$, which can be shown to bring negligeable contributions.
In our kinematical range for this study $\xi \sim 0.15$ and $x \sim {10}^{-3}$, that is $v \sim 10$, ${\bar{\omega}} \sim 0.4,$ we can use the assymptotic expansion for the Bessel functions (at large $v$)
\beq
{I}_{\nu}(v) \simeq \frac{\exp (v)}{{2 {\pi} v}^{\frac{1}{2}}}\left[1-\frac{1}{2v}\frac{\Gamma(\nu+\frac{3}{2})}{\Gamma(\nu-\frac{1}{2})}\right]
\eq
which gives 
\beq
\frac{{\partial \ln} xF_s } {{\partial \ln}{1 \over {x}}}
\sim
\bar{\omega} \frac{I_3(v)}{I_2(v)}\left[
1-\frac{b_1 {\bar{\omega}}}{v}+O(\frac{{{\bar{\omega}}}^{2}}{v})\right]
\eq
So the absolute correction due to the term $b_1$ is of the order of
$b_1 \frac{{{\bar{\omega}}}^{2}}{v}$ which is negligeable in the considered kinematical range. The terms $b_{i}, i>1$ are even more negligeable.
We deduce that $\frac{{\partial \ln} xF_s } {{\partial \ln}{1 \over {x}}}$ does not depend on the different $b_i'$ s.
Hence the GRV type parametrizations ought to verify the following prediction 
\beq
\frac{{\partial \ln} xF_s } {{\partial \ln}{1 \over {x}}} \simeq
\bar{\omega} \frac{I_3(v)}{I_2(v)}
\eq
which shows that this observable does not depend on the valence like input distribution.
Using the precedent assymptotic expansion of the Bessel functions we can find an   expansion (at large $v$)
\beq
\frac{I_3(v)}{I_2(v)} \simeq 1-\frac{5}{2v}
\eq
which leads to 
\beq
\frac{{\partial \ln} xF_s } {{\partial \ln}{1 \over {x}}} \simeq
{(\frac{4 N_c \xi}{ \ln 1/x})}^{\frac{1}{2}}
-\frac{5}{4 \ln 1/x}.
\eq
Hence 
\beq
\frac {\partial^2}{\left(\partial \ln Q^2\right)^2} \left(  
\frac {\partial \ln xF_s}{\partial \ln {1 \over {x}}} \right)
\sim \frac{{\partial^3 \ln} xF_s } {{\partial \ln}{1 \over {x}}{\partial^2 \xi}} < 0
\eq
This gives us an information on the concavity of the function $\lambda(Q^2)$ in the DGLAP scheme with the rightmost singularity imposed by the DGLAP kernel.
\\

\no
{\bf b} -In contrast to GRV type parametrizations, the LBY parametrization \cite{LBY} uses an input $\omega-$plane singularity fixed in $Q^2$ and located at the right of $\omega=0$. Thus, our previous derivation  does not apply in this case. Starting from the LBY formulation of the singlet structure function
\beq
\label{LBY}
<e^2> xF_s(x,Q^2) = [B_s(Q^2) x^{-\lambda_s} + C_s(Q^2)] (1-x)^{\nu(Q^2)}
\eq
where $<e^2>$ is the average charge for $N_f$ flavours and $B_s$, $C_s$ and $\nu$ are $Q^2$-dependent functions \cite{LBY}. $\lambda_s >0$ defines the location of the rightmost $\omega-$plane singularity and is $Q^2-$independent, but for charm and bottom thresholds.  
At small $x$,
\beq
\label{predLBY}
\frac{\partial \ln xF_s } {\partial \ln \frac{1}{x}} \simeq \lambda_s,
\eq
up to the correction due to  the phenomenological factor $C_s(Q^2).$
\\
\no
{\bf c} -In the framework of the QCD dipole model, the BFKL dynamics also provides predictions for $\frac{\partial \ln xF_s } {\partial \ln \frac{1}{x}}$. It gives \cite{npr,nprw}
\beq
\label{predF2}
F_s = C a^{1/2} x^{-\alpha_{P}} \frac{Q}{Q_0} e^{- \frac{a}{2} \ln^2 \frac{Q}{Q_0}}
\eq
where
\beq
\label{a}
\alpha_{P}  = 1 + \frac{4 \bar{\alpha} N_{C} \ln 2}{\pi} \quad {\rm and} \quad a  = \left(\frac{\bar{\alpha} N_c}{\pi} 7 \zeta(3) \ln\frac{1}{x}\right) ^{-1}
\eq
$\alpha_P$ is the well-known BFKL Pomeron intercept which is a constant, since the strong coupling constant $\bar{\alpha}$ is held fixed in this scheme.
$C$ and $Q_0$ are non-perturbative parameters to be determined by the fit.
 We get
\beq
\label{obserbfkl}
\frac{\partial \ln F_s } {\partial \ln \frac{1}{x}}
= \alpha_p - \de \frac{1}{\ln \frac{1}{x}} + \frac{1}{14 \frac{\bar{\alpha} N_c}{\pi} \zeta (3) \ln ^2\frac{1}{x}} \ln^2 \frac{Q}{Q_0}
\eq
\no
At fixed $Q^2$ and $x \rightarrow 0$, one recovers the usual BFKL Pomeron intercept. 
Note that a $Q^2$-dependence stems from the last term in equation (\ref{obserbfkl}). Here we have, contrary to the  GRV scheme 
\beq
\frac{{\partial^3 \ln} xF_s } {{\partial \ln}{1 \over {x}}{\partial^2 \xi}} > 0
\eq
\\

\no
III - Let us now discuss the phenomenological consequences of the previous calculations.
In the following discussion, we have to take into account the
difference, due to the non-singlet contributions, between $xF_s$ and $<e^2>^{-1}F_2$, where $<e^2> = 2/9 (5/18)$ for $N_f =3 (4).$
However the non-singlet contribution is expected to be regular at $\omega=0$. Its QCD evolution receives no contribution from the gluon. Thus, one can apply our study of $ \frac{\partial \ln F}{\partial \ln \frac{1}{x}} $ to $F_2$ as well. 
In practice, our phenomenological discussion does include the non-singlet contribution when discussing the GRV and LBY parametrizations.
Anyway, we have verified that this 
component is rather weak at small-$x.$ 

\no
At this stage, a comment is in order about the non-perturbative value of $\mu^2$. Indeed, in the leading-order version of the GRV parametrization, the evolution variable $\xi(\mu^2)$ (see formula (\ref{xidef})) is defined using the parameter values $N_f = 3$, $\mu = 480\ MeV$ 
and $\lqcd = 232\ MeV$.
We follow here an argument similar to the one given in ref.\cite{wil}. Starting from a perturbative scale $Q_0^2$ and evolving the structure function up to $Q^2$, the renormalization group (DGLAP) predicts
\beq
\label{regul}
xF_s \propto K^{{\frac{4 N_c} {\omega}}} \ e^{{\frac{4 N_c} {\omega}} 
\xi(Q_0^2)} (\omega f(\omega))
\eq
where $\xi(Q_0^2)$ is defined as in equation (\ref{xidef})
\beq
\label{xi}
\xi(Q_0^2) = {\frac{1}{11 - {\frac{2}{3}} N_f}} \ln \left(\frac{{\ln}Q^2/\Lambda^2}{\ln  {Q_o}^2/\Lambda^2} \right).
\eq
Note that ${\frac{4 N_c} {\omega}}$ is the singular part of the DGLAP anomalous dimension $\nu_+$ at $\omega=0$. The form of the prefactor $K^{{\frac{4 N_c} {\omega}}}$ ensures that the physical result does not depend on the arbitrariness of the factorization scale $Q_0^2$ in the {\it perturbative} range. The constant
$K$ cannot be predicted but has to be larger than one in order to get positive structure functions. using the functional form (\ref{xidef}),
one can define $\mu^2$ in such a way that
\beq
\label{redefxi}
\xi(\mu^2) = \xi(Q_0^2) + \ln K
\eq
where $\xi$ is defined in equation (\ref{xi}).
Since $K > 1$, one expect $\mu < Q_o$. Eventually, $\mu^2$ will be in the non-perturbative domain. This is
indeed the case in the GRV parametrization,  which uses an {\it effective} non-perturbative parameter $\mu^2$  and reproduces the data. In conclusion, $\mu$ is an effective scale for the renormalization group evolution in the whole  perturbative range $Q >> \mu$ without requiring the validity of the renormalization group in the non-perturbative range $Q \sim \mu$. Note that a
similar argument also holds when applying the QCD dipole model for proton structure functions \cite{nprw}. \\
\no
In fig.1 we compare the results of the GRV parametrization for $ \frac{\partial \ln xF}{\partial \ln \frac{1}{x}} $ with the prediction 
\beq
\label{pred}
\frac{{\partial \ln} xF_s } {{\partial \ln}{1 \over {x}}} \simeq
\bar{\omega} \frac{I_3(v)}{I_2(v)}
\eq   
This comparison is displayed in fig.1a as a fonction of $x$ for four different values of $Q^2$ $(5, 20, 100, 800 \ GeV^2)$. The agreement with the
prediction (\ref{pred}) is reasonable for $10^{-4} < x < 2.10 ^{-3}$. Note that since this observable is a smooth function of $x$, the slope is extracted
from a global fit of each $Q^2$ bin by a function of the type $x^{-\lambda}$. We will come back 
to this point later.
In fig.1b, we have plotted the  $Q^2$-dependence of $ \frac{\partial \ln xF}{\partial \ln \frac{1}{x}} $ for $x=10^{-3}$  and $10^{-4}$. In both figures, a slight difference is observed between the two determinations. This might be due to the fact that the GRV parametrization is not an exact solution of the moment equations. A better agreement would be obtained with a slight change of $\mu^2$ from $.23$ to $.24 GeV^2$.
 
\no
Fig.1a and fig.1b illustrate our claim that the evolution of the considered observable is indeed dominated by the behaviour of the leading
anomalous dimension.

\no
In fig.2, we display the data on $\lambda$ defined in formula (\ref{lambda}) and the prediction of the different parametrizations and evolution equations.
The procedure for comparing models to data has been to use the same averaging
in both cases. For each value of $Q^2$, we have used the value of $\lambda$ given by a fit of the form $x^{-\lambda}$ for $10^{-4} \leq x \leq 10^{-2}$. 
Note that we have performed our own fit for E665 data \cite{e665} while we used the published H1 data for $\lambda$ with $x \leq 10^{-1}$. We have checked that restricting the fit to the range $10^{-4} \leq x \leq 10^{-2}$ does not change the result. \no
Fig.2 shows that the $\lambda$ value obtained from the GRV parametrization is well reproduced by our prediction. It thus exhibits the universality property of $\lambda$ at small $x$ in the whole $Q^2$ range, since this universal value $\lambda$ is determined with essentially only one parameter
$\mu^2$. Hence, a more accurate measurement of this observable may disentangle the nagging problem of the existence of the singular nature of the input.

\no
The different parametrizations are compatible with the data. 
The DGLAP (LBY) parametrization is  displayed for $Q^2 > 10 \ GeV^2$ in its domain of validity.
We note in this case that $\lambda$ is slowly varying with $Q^2$ and lower than the  $\lambda_s = .36$ in formula (\ref{predLBY}). This might be due to the extra singlet factor $C_s$ in equation (\ref{LBY}).

\no
The BFKL (dipole) prediction is  satisfactory. It is interesting to note that it predicts a $Q^2$ variation of $\lambda$, showing that a model satisfying the BFKL dynamics combined with $k_T$-factorization induces a $Q^2$ variation of the slope, even if the $\omega-$singularity is fixed. We note a difference with the DGLAP predictions in the small $Q^2$ range $(1 \leq Q^2 \leq 10 \ GeV^2)$. Accurate data in this range might distinguish 
between the two different approaches.

In this context, we have also included in fig.2 the very recent H1 1995 preliminary results for $\lambda$ \cite{oups}. They unravel a tendency in favour of the DGLAP(kernel) scenario. This new measurement, if confirmed and made more
precise in the future, shows the reliability of measuring $\lambda$ for 
distinguishing the QCD origin of the rise of $F_2.$
The  quality of the experimental determination deserves more study at the theoretical level, for instance a more precise determination of the QCD dipole predictions at 
low $Q^2,$ a study of the flavour threshold effects in the LBY approach and 
a DGLAP kernel considered at higher perturbation order. Work is in progress in those directions.

One could wonder whether the observable $ \frac{\partial \ln xF_s}{\partial \ln Q^2}$ (or$ \frac{\partial \ln xF_s}{\partial \xi}$) obeys the same
universality property as $ \frac{\partial \ln xF_s}{\partial \ln \frac{1}{x}}$ in GRV type models, since the DGLAP kernel is the dominant singularity in the $\omega-$plane. 
The following study shows that this is not the case. It turns out that it depends strongly on the input,
and it is  thus not dominated by the anomalous dimension.
A developpement similar to the previous one for the determination of $\frac{{\partial \ln} xF_s}{{\partial \ln} 1/x}$ leads to 

\beq
\frac{\partial \ln xF_s}{\partial \xi}=
\frac{4N_c}{\bar{\omega}} \frac{I_1(v)+\Sigma_i b_i {\bar{\omega}}^{i} I_{i+1} (v)}
{I_2(v)+\Sigma_i b_i {\bar{\omega}}^{i} I_{i+2} (v)}
\eq
with $\nu_{+}(\omega) \simeq \frac{4N_c}{\omega}$. 
\no
Now, if we take  an accurate developpement for $\nu_+$, see (7), we get 
\beq
\frac{\partial \ln xF_s}{\partial \xi}=
-a+\frac{4N_c}{\bar{\omega}} \frac{I_1(v)+\Sigma_{i}  b_i {\bar{\omega}}^{i} I_{i+1} (v)}
{I_2(v)+\Sigma_i b_i {\bar{\omega}}^{i} I_{i+2} (v)}
\eq
The assymptotic expansion for the Bessel functions leads to
\beq
\frac{\partial \ln xF_s}{\partial \xi}=
-a+\frac{4N_c}{\bar{\omega}} \frac{I_1(v))}
{I_2(v)}[1-\frac{b_1{\bar{\omega}}}{v}]
\eq
\no
Note that the main contribution to $\frac{\partial \ln xF_s}{\partial \xi}$ behaves as $\frac{1}{\bar{\omega}}$ in contradistinction with the previous case where $\frac{\partial \ln xF_s}{\partial \ln 1/x}$ is of  order ${\bar{\omega}}$. This is the origin of the non universality of this observable as shown below.

\no
Indeed the term $b_1$ contributes to $\frac{\partial \ln xF_s}{\partial \xi}$ with a coefficient $\frac{4N_c}{v}$ to be compared with its contribution to $\frac{\partial \ln xF_s}{\partial \ln 1/x}$ where the coefficient is 
$\frac{{{\bar{\omega}}}^{2}}{v}$. As ${\bar{\omega}} \sim 0.4$, $v \sim 10$, $4N_c=12$, the correction in $b_1$ to the kernel contribution 
  
\beq
\frac{\partial \ln xF_s}{\partial \xi} \sim
-a+\frac{4N_c}{\bar{\omega}} \frac{I_1(v)}
{I_2(v)}
\eq
\no
is important, so this observable is very sensitive to the corrections due to the terms $b_i$. Moreover $b_1\frac{4N_c}{v} \sim a$, so this derivative is sensitive to an higher order of developpement of $\nu_{+}(\omega)$ in $\omega$.
We can undestand this point by writing exactly the derivative with respect to $\xi$
\beq
\frac{\partial \ln xF_s}{\partial \xi}={1 \over 2i\pi} \int^{\omega_0 + i \infty}_{\omega_0 - i \infty} 
\omega f(\omega) \nu_{+}(\omega){\rm e}^{\  \omega {\ln}1/x} {\rm e}^{\frac{4 N_c} {\omega} \xi}  {\rm d} \omega,
\eq
\no
The terms coming from the development of $\nu_{+}(\omega)$ will be mixed with those from $f(\omega)$ and we can not separate both contributions as it is the case for $\frac{\partial \ln xF_s}{\partial \ln 1/x}$.
Thus $\frac{\partial \ln xF_s}{\partial \xi}$ does not obey the same universality property as $\frac{\partial \ln xF_s}{\partial \ln 1/x}$. We can notice here that another explanation using a saddle point method introduced in ref.\cite{npw}  leads by a different method to the same conclusion \cite{these}. 
\\

To summarize, we have shown that, provided that the non-perturbative input
is regular at the right of the $\omega=0$  singularity of the dominant DGLAP anomalous dimension, the 
observable $\lambda = \left<\frac{\partial \ln xF_2}{\partial \ln \frac{1}{x}} \right>$ is input-independent in the perturbative $Q^2$ regime at small $x$.
Other models, namely BFKL(dipole), DGLAP(LBY) give different values for $\lambda$, which are compatible 
with present published data. This is an incentive for the experimentalists to get a better accuracy, and for the theoreticians to refine the predictions in order  
to distinguish these different QCD interpretations in the near future. 
\\

\no
{\bf Acknowledgements:} 

\no
We are very thankful to Joel Feltesse for fruitful discussions.
\eject

\eject
{\bf FIGURE CAPTIONS}

{\bf Figure 1a}
Comparison of the GRV parametrization for $\frac{{\partial \ln} xF_s } {{\partial \ln}{1 \over {x}}}$ with our prediction $\bar{\omega} \frac{I_3(v)}{I_2(v)}$ for the same observable (called DGLAP(Kernel) on the figure).   This comparison is displayed  as a fonction of $x$ for 4 different values of $Q^2$ $(5, 20, 100, 800 \ GeV^2)$.
\\

\no
{\bf Figure 1b}
Comparison of the GRV parametrization for $\frac{{\partial \ln} xF_s } {{\partial \ln}{1 \over {x}}}$ with our prediction $\bar{\omega} \frac{I_3(v)}{I_2(v)}$ for the same observable (called DGLAP(Kernel) on the figure). This comparison is displayed  as a fonction of 
$Q^2$ for two different values of $x$ laying in the kinematical range of our study : $x={10}^{-3},  x={10}^{-4}$.
\\

\no
{\bf Figure 2}
Display of the data on $\lambda$ (H1: \cite{h1}, \cite{oups} and E665: \cite{e665}) compared with the prediction of the different parametrizations and evolution equations.

\eject
\input epsf
\vsize=30.truecm
\hsize=16.truecm

\epsfxsize=16.cm{\centerline{\epsfbox{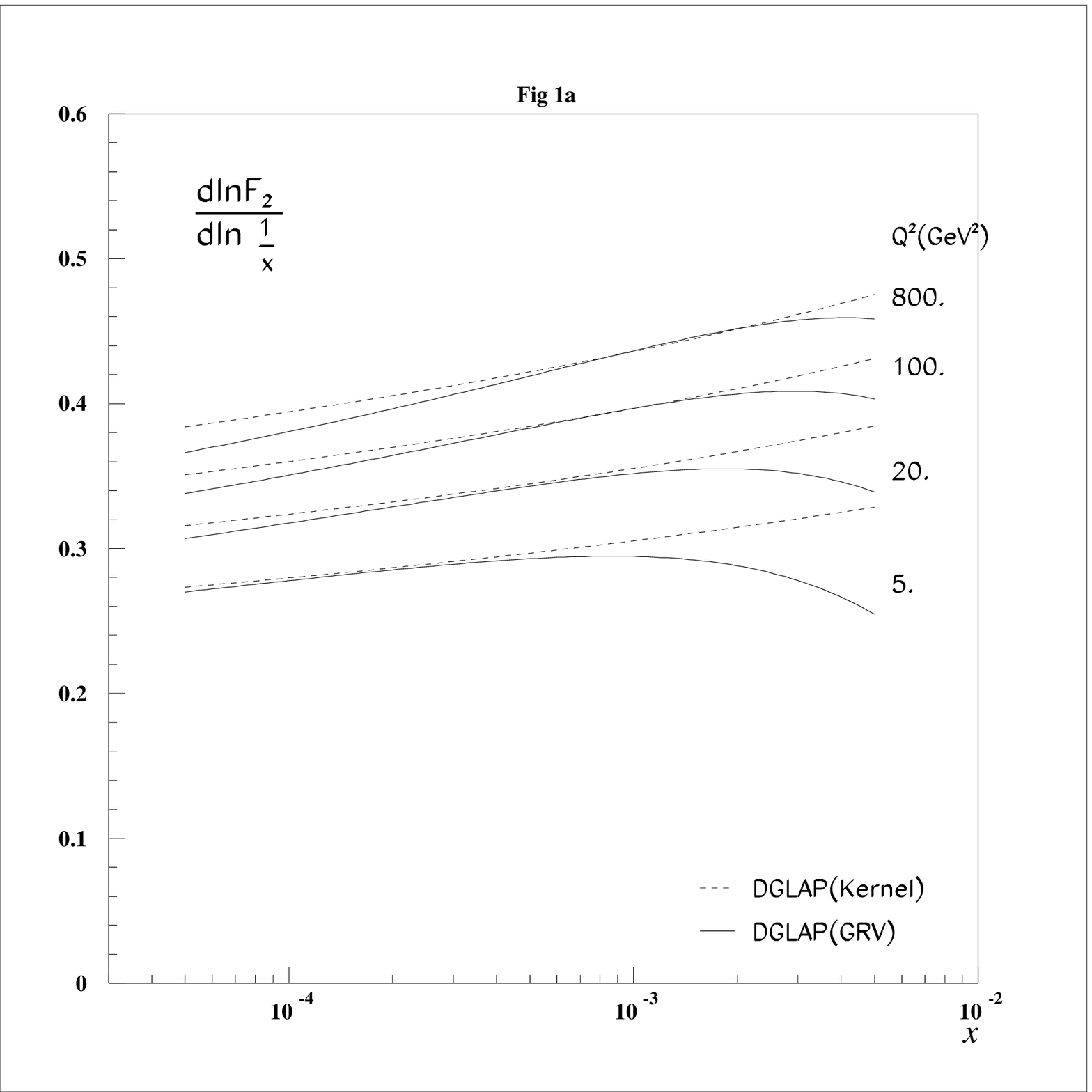}}}
\eject

\vsize=30.truecm
\hsize=16.truecm
\epsfxsize=16.cm{\centerline{\epsfbox{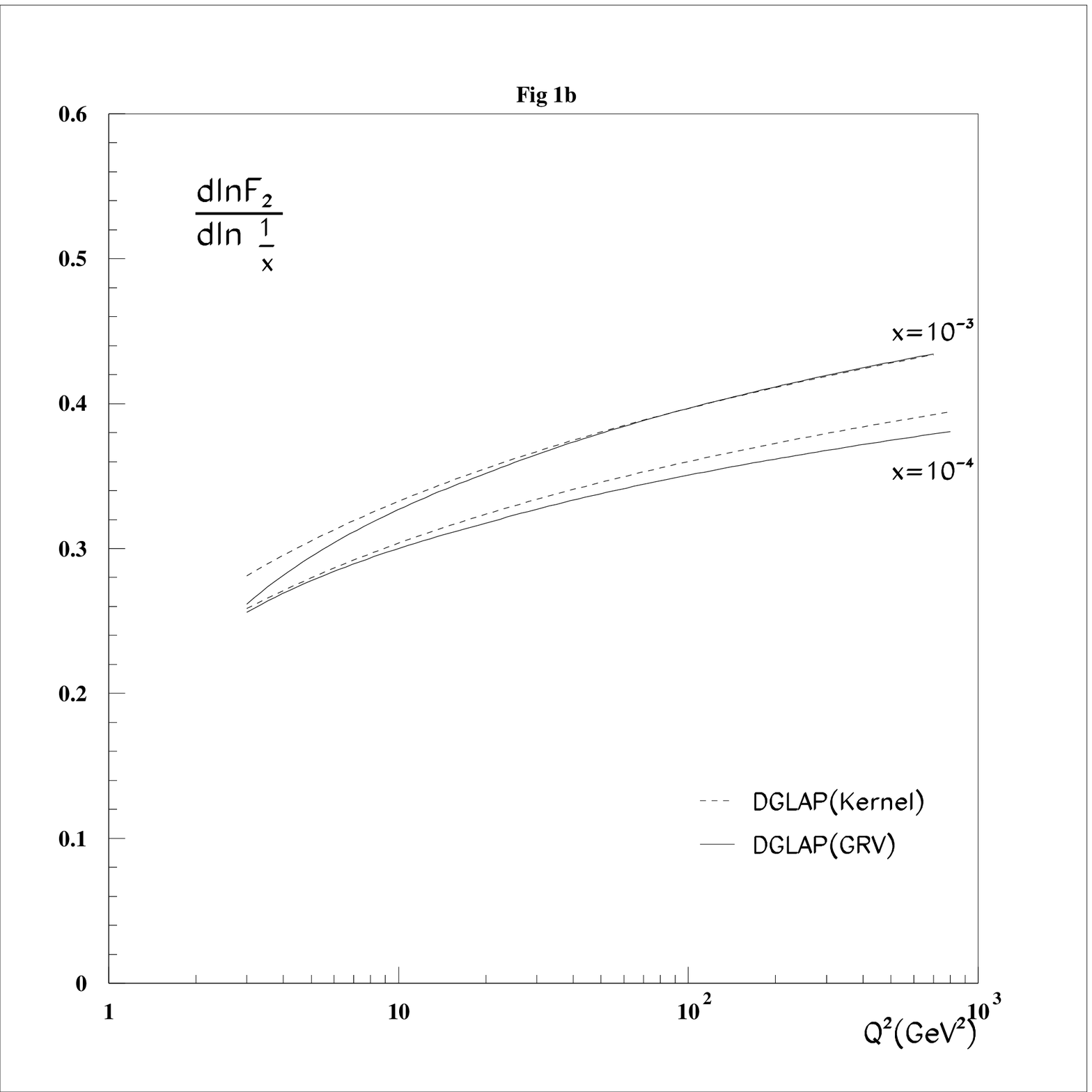}}}
\eject

\vsize=30.truecm
\hsize=16.truecm
\epsfxsize=16.cm{\centerline{\epsfbox{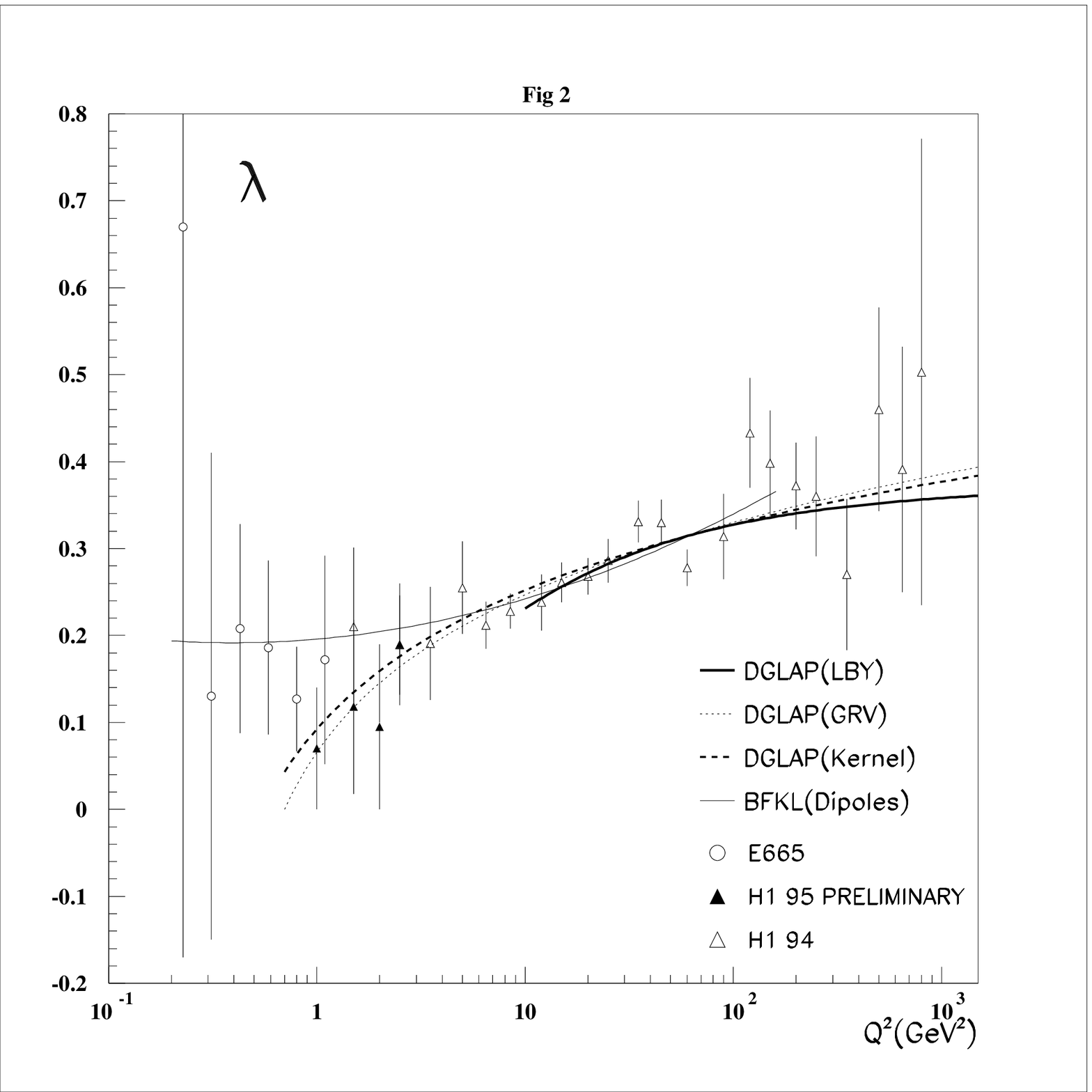}}}
\eject

\end{document}